\documentclass[runningheads]{llncs}

\usepackage{url}
\usepackage{amsmath}
\usepackage{graphicx}
\usepackage{multirow}

\begin{document}

\title{Inner speech recognition through electroencephalographic signals}

\author{Francesca Gasparini\inst{1,2}\orcidID{0000-0002-6279-6660} \and
	Elisa Cazzaniga\inst{1} \and
    Aurora Saibene\inst{1,2}\orcidID{0000-0002-4405-8234}
}
\authorrunning{F. Gasparini et al.}

\institute{University of Milano-Bicocca, Viale Sarca 336, 20126, Milano, Italy \\
	\email{aurora.saibene@unimib.it, e.cazzaniga@campus.unimib.it, francesca.gasparini@unimib.it}
	\and
	NeuroMI, Milan Center for Neuroscience, Piazza dell’Ateneo Nuovo 1, 20126, Milano, Italy 
}

\maketitle

\begin{abstract}
This work focuses on inner speech recognition starting from EEG signals. Inner speech recognition is defined as the internalized process in which the person thinks in pure meanings, generally associated with an auditory imagery of own inner “voice”.
The decoding of the EEG into text should be understood as the classification of a limited number of words (commands) or the presence of phonemes (units of sound that make up words). Speech-related BCIs provide effective vocal communication strategies for controlling devices through speech commands interpreted from brain signals, improving the quality of life of people who have lost the capability to speak, by restoring communication with their environment.
Two public inner speech datasets are analysed. Using this data, some classification models are studied and implemented starting from basic methods such as Support Vector Machines, to ensemble methods such as the eXtreme Gradient Boosting classifier up to the use of neural networks such as Long Short Term Memory (LSTM) and Bidirectional Long Short Term Memory (BiLSTM). With the LSTM and BiLSTM models, generally not used in the literature of inner speech recognition, results in line with or superior to those present in the state-of-the-art are obtained. 
	
\end{abstract}

\section{Introduction} \label{sec:introduction}
Human speech production is a complex motor process that starts in the brain and ends with respiratory, laryngeal, and articulatory gestures for creating acoustic signals of verbal communication. Physiological measurements using specialized sensors and methods can be made at each level of speech processing, including the central and peripheral nervous systems, muscular action potentials, speech kinematics (tongue, lips, jaw), and sound pressure~\cite{Schultz2017BiosignalBasedSC}.
However, there are cases of subjects suffering from neurodegenerative diseases or motor disorders that prevent the normal activity of signal transmission from the brain to the peripheral areas. These subjects are prevented from communicating or carrying out certain actions.

Brain Computer Interfaces (BCIs) are promising technologies for improving the quality of life of people who have lost the capability to move or speak, by restoring communication with their environment. 
A BCI is a system that makes possible the interaction between an individual and a computer without using the brain’s normal output pathways of peripheral nerves and muscles.
In particular, speech-related BCI technologies provide neuro-prosthetic help for people with speaking disabilities, neuro-muscular disorders and diseases. It can equip these users with a medium to communicate and express their thoughts, thereby improving the quality of rehabilitation and clinical neurology~\cite{saha2019speak}. 
Speech-related paradigms, based on either silent, imagined or inner speech provide a more natural way for controlling external devices~\cite{nieto2022thinking}. 

There are different types of brain-signal recording techniques that are mainly divided into invasive or non-invasive methods. The first ones involve implanting electrodes directly into the brain. They provide better spatial and temporal resolution, also increasing the quality of the signal obtained. However, invasive technologies have problems related to usability and the need for surgical intervention on the subject.
This is why non-invasive techniques are increasingly used in BCI research.
Among the non-invasive technologies, the electroencephalogram (EEG) is the most used method for measuring the electrical activity of the brain from the human scalp. It has an exceedingly high time resolution, it is simple to record and it is sufficiently inexpensive~\cite{gu2021eeg}. Over the years, EEG hardware technology has evolved and several wireless multichannel systems have emerged that deliver high quality EEG and physiological signals in a simpler, more convenient and comfortable design than the traditional, cumbersome systems.

This paper focuses on inner speech recognition starting from EEG signals, where the basic definition of \textit{inner speech} is~\cite{alderson2015inner} "the subjective experience of language in the absence of overt and audible articulation".

As suggested in \cite{van2021inner}, there is evidence from past neuroscience research that inner speech engages brain regions that are commonly associated with language comprehension and production~\cite{amit2017asymmetrical}. This includes temporal, frontal and sensorimotor areas predominantly in the left hemisphere of the brain~\cite{amit2017asymmetrical} \cite{bocquelet2016key}. Therefore, by monitoring these brain areas, it is theoretically possible to develop an inner speech BCI that classifies neural representations of imagined word~\cite{bocquelet2016key}. 

In section \ref{sec:related-works} the studies in the field of inner speech are described. Section \ref{sec:datasets} presents the two publicly available datasets used for the analyses proposed in section  \ref{sec:proposed-approaches}. In section \ref{sec:results-and-discussion} the results obtained with our models are presented and discussed. Finally, in section \ref{sec:conclusion} some conclusions are proposed.

\section{Related works}\label{sec:related-works}

Most studies on classification of inner speech focus on invasive methods such as electrocorticography (ECoG) \cite{martin2018decoding} as they provide higher spatial resolution while fewer studies concerning inner speech classification using EEG data are available \cite{panachakel2021decoding}. It is important for a BCI application to be non-invasive, accessible and easy to implement so that it can be used by a large number of subjects.

Inner speech recognition is generally faced considering phonemes, in general vowels or syllables, such as /ba/ or /ku/, or simple words such as left, right, up and down, in subject-dependent approaches.

Preliminary works were conducted with very few participants and syllables by D’Zmura et al. \cite{d2009toward}, where EEG waveform envelopes have been adopted to recognize EEG patterns.
Also Brigham and Kumar \cite{brigham2010imagined} 
and Deng et al. \cite{deng2010eeg} 
considered the recognition of two syllables. 
In the first work, the accuracy obtained for the 7 subjects ranges from 46\% to 88\%. They preprocessed raw EEG data to reduce the
effects of artifacts and noise, and applied k-Nearest Neighbor classifier to autoregressive coefficients extracted as features. While Deng end colleagues using Hilbert spectra and linear discriminant analysis  recognized the two syllables imagined in three different rhythms, for a 6 classes task,  with accuracy ranging from 19\% to 22\%. 

Considering works where recognition of phonemes have been investigated, Da Salla et.al \cite{dasalla2009single} and \cite{dasalla2009spatial}, analyzed the recognition of three tasks: /a/, /u/, and rest, obtaining from 68\% to 79\% of accuracy by using common spatial patterns. On the same dataset, several other researchers have been tested different models obtaining promising results \cite{idrees2016vowel}, \cite{riaz2014inter}.

Kim et al. \cite{kim2014eeg}, instead, consider three vowels /a/, /i/ and /u/ and applied multivariate empirical mode decomposition and common
spatial pattern for feature extraction together with linear discriminant analysis, reaching around 70 \% of accuracy. 


Few representative studies that try to recognize imagined words using EEG data are reported in the literature. 
Given the complexity of the task, the number of terms considered is generally limited.\\
Suppes et al. \cite{suppes1997brain} proposed an experiment in which five subjects performed the internal speech considering the following words: first, second, third, yes, no, right, and left for all subjects with the addition of to, too and hear for the last three subjects.\\
In the work performed by Wang at al. \cite{wang2013analysis}, eight Chinese subjects were required to read in mind two Chinese characters (that meant left and one). They were able to distinguish between the two characters and the rest state. Feature vectors of EEG signals were extracted using CSP, and then these vectors were classified with SVM. Accuracies between 73.65\% and 95.76\% were obtained when comparing between each of the imagined words and the rest state. A mean accuracy of 82.3\% was achieved between the two words themselves.\\
Salama et al. \cite{salama2014recognition} implemented different types of classifiers such as SVM, discriminant analysis, self-organizing map, feed-forward back-propagation and a combination of them, to recognize two words (Yes and No). They used a single electrode EEG device to collect data from seven subjects and the accuracy obtained ranges from 57\% to 59\%.\\
In \cite{mohanchandra2016communication}, Mohanchandra at al. constructed a one-against-all multiclass SVM classifier to discriminate five subvocalized words (water, help, thanks, food, and stop) and reported an accuracy ranging from 60\% to 92\%. \\
In the González-Castañeda at al. \cite{gonzalez2017sonification} analyses, some techniques of sonification and textification are applied, which allows to characterize EEG signals as either an audio signal or a text document. Five imagined words (up, down, left, right) and 27 subjects are considered. The average accuracy rate using the EEG textified signals were 83.34\%.\\
Using the data from six subjects, \cite{nguyen2017inferring} reported an average accuracy of 50.1\% for the three-short words (in, out and up) classification problem and 66.2\% for the two long words classification problem (cooperate and independent), using a Multi-Class Relevance Vector Machine (MRVM). In order to evaluate the effect of the sound, three phonemes were used, namely /a/, /i/ and /u/ obtaining an accuracy of 49.0\%.
Coretto et al. 
\cite{coretto2017open} that collected one of the two datasets considered in this paper, reported a mean recognition rate of 22.32\% in classifying five Spanish vowels and 18.58\% in classifying six Spanish words using a Random Forest (RF) algorithm.
Using the same dataset, in \cite{cooney2020evaluation} 30.00\% and 24.97\% accuracies are obtained respectively for vowels and words using CNNs.\\
Recently, Bram van den Berg et al. \cite{van2021inner}, working on the \textit{Thinking Out Loud} dataset \cite{nieto2022thinking}, also considered in our work, reported an average accuracy of 29.7\% for a four-word classification task using a 2D CNN based on the EEGNet architecture \cite{lawhern2018eegnet}.

\section{Datasets}\label{sec:datasets}
The testing of the proposed strategies is performed on two publicly available datasets, i.e., the \textit{Thinking Out Loud}~\cite{nieto2022thinking} and the \textit{Imagined Speech}~\cite{coretto2017open} datasets. In particular, the last dataset is used to check the validity of the resulting best approach for the \textit{Thinking Out Loud} one.

\subsection{Thinking Out Loud dataset}\label{sec:thinkingoutloud}
The first literature dataset chosen to conduct the subsequent analyses is the \textit{Thinking Out Loud}~\cite{nieto2022thinking} one, which is focused on an inner speech paradigm intended for the control of a BCI system through imagination of Spanish words.\\ 
The selected Spanish words are \textit{arriba} (up), \textit{abajo} (down), \textit{derecha} (right), and \textit{izquierda} (left). Notice that the words were presented randomly with a visual cue.\\
Ten (four females) healthy right-handed subjects with mean $\pm$ std age 34 $\pm$ 10, without any hearing or speech loss, nor any previous BCI experience, participated in the experiment, which consisted of three experimental conditions, i.e., inner/pronounced speech and visualized condition.
During the \textit{inner speech} condition the participant was asked to imagine his/her own voice, repeating the corresponding word. Instead, the participant was asked to repeatedly pronounce aloud the word corresponding to each visual cue during the \textit{pronounced speech} condition. Finally, the \textit{visualized condition} corresponded to the tasks during which the participant was asked to focus on mentally moving a circle shown in the center of a screen in the direction indicated by a visual cue.

\begin{figure}[h]
	\centering
	\includegraphics[width=0.9\linewidth]{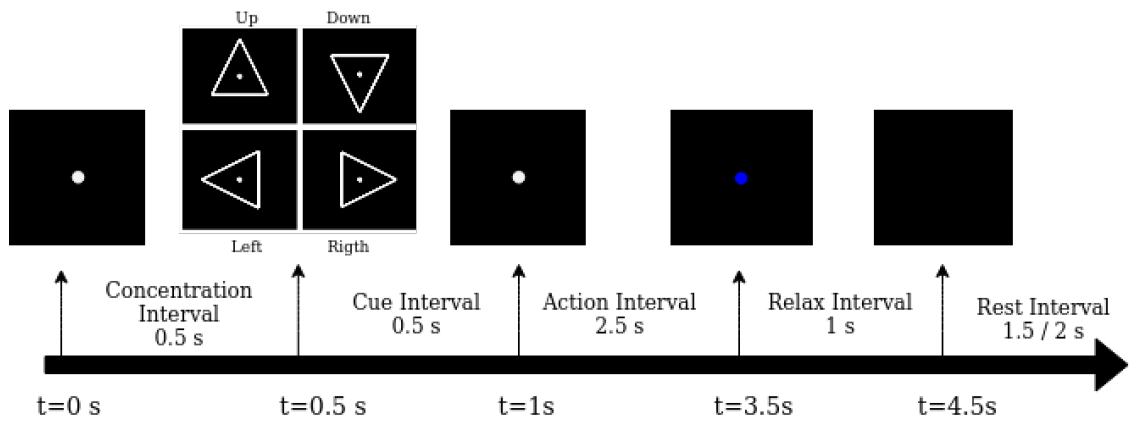}
	\caption{Trial workflow taken from the \textit{Thinking Out Loud} dataset original paper~\cite{nieto2022thinking}.}
	\label{fig:experim-proced}
\end{figure}

Each subject participated in 3 consecutive sessions (200 words/session), separated by a break. Each session consisted of a baseline recording (15s), the \textit{pronounced speech} run, two \textit{inner speech} runs, and two \textit{visualized condition} runs. Each run was constituted by a series of trials containing the different experimental tasks. \\ 
Figure \ref{fig:experim-proced} shows the organization of each \textit{inner speech} trial, under investigation in this paper. A white circle was shown in the center of the screen and the subject was asked to stare at it without blinking. Subsequently, a white triangle was shown pointing in one of the four directions corresponding to the chosen Spanish words. When the triangle disappeared and the white circle was presented again, the subject had to perform the indicated task. The task execution had to be stopped when the white circle turned blue. The subject has been asked to control eye blinking until the circle disappeared. Finally, to evaluate the participants' attention, the subjects were asked to indicate the last inner speech and visualized conditions after a random number of trials. The subject answered using keyboard arrows and feedback was displayed.


The data acquisition was performed using 128 active EEG wet electrodes and 8 external active EOG/EMG wet electrodes. The resolution was of 24 bits resolution and 1024 Hz sampling rate applied.

%


The EEG signals of the \textit{Thinking Out Loud}
dataset were preprocessed by its authors. The preprocessing included a band pass filter between 0.5-100 Hz, a notch filter at 50 Hz and downsampling to 254 Hz. 

Please, find further details on the original dataset paper~\cite{nieto2022thinking}.

\subsection{Imagined Speech dataset} \label{sec:imagined-speech-dataset}
The \textit{Imagine Speech} dataset~\cite{coretto2017open} was chosen to confirm the validity of the model obtaining the best results on the \textit{Thinking Out Loud} dataset. In fact, similar experimental conditions are presented considering Spanish words and also vowels. \\
In fact, the vowels /a/, /e/, /i /, /o/ and /u/ have been selected due to their acoustic stationarity, simplicity and lack of meaning by themselves. While the Spanish words \textit{arriba} (up), \textit{abajo} (down), \textit{derecha} (right), \textit{izquierda} (left), \textit{adelante} (forward), and \textit{atras} (backward) were chosen as possible BCI commands to control the movements of an external device.

\begin{figure}[!ht]
	\centering
	\includegraphics[width=0.9\linewidth]{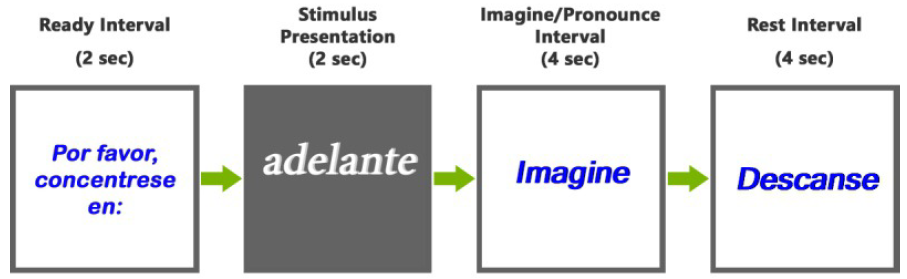}
	\caption{Sequence time course for the presentation of one stimulus, in this particular case for the word \textit{adelante} and under the \textit{imagined speech} condition. Graphic taken from the original dataset paper~\cite{coretto2017open}.}
	\label{fig:imagined-speech-dataset}
\end{figure}

Fifteen (seven females) healthy subjects with mean age of 25 years old, without any hearing or speech loss, participated in the experiment.
Only one of the subjects reported to be left-handed, while the rest were right-handed.\\
EEG signals were recorded under two conditions: \textit{imagined speech} and \textit{pronounced speech}. During \textit{imagined speech}, the subjects had to imagine pronouncing the word without moving muscles or producing sounds.\\
Target stimuli were presented in a sequence comprised of four intervals of predefined duration (Figure \ref{fig:imagined-speech-dataset}). During the ready interval (2 s), the subject is informed that the rest interval finished and a new cue would be displayed soon. Afterwards, the target word is presented, both visually and acoustically, during the stimulus presentation interval (2 s). In the Imagine/Pronounce stage an image displays the task requested (either imagined or pronounced speech). It is in this stage that the subject has to imagine the pronunciation or pronounce the word given as a cue. In the case that the word is a vowel, the subject must perform the task during the complete 4 seconds of this interval duration, while if the word is a command, a sequence of three audible clicks will indicate when to imagine or pronounce the target word. Finally, during the rest interval (4 s) the subject is allowed to move, swallow or blink. 

The number of trials performed by each subject for each word is fifty, with forty corresponding to the \textit{imagined speech} condition and the other ten belonging to the \textit{pronounced speech} one.

EEG signals were recorded using Ag-AgCl cup electrodes, attached to the scalp according to the 10-20 international system and with conductive paste. No electrode cap was used. F3, F4, C3, C4, P3, and P4 were chosen as active electrodes, while reference and ground electrodes were placed on the left and right mastoids. 
The EEG signals were acquired with 1024 Hz sampling rate and 16 bit resolution. 


The EEG signals of the \textit{Imagined Speech} dataset were preprocessed by its authors. The preprocessing included a band pass filter between 2-40 Hz.

Please, find further details on the original dataset paper~\cite{coretto2017open}.

\section{Proposed approaches}\label{sec:proposed-approaches}

Some classification models are studied and implemented starting from basic methods such as SVM, to ensemble methods such as the XGBoost classifier up to the use of neural networks such as LSTM and BiLSTM.

\subsection{Machine Learning approaches}
Since inner speech recognition is a very complicated task, a simpler preliminary analysis was performed, which consists of binary classification between the resting state and the action interval. The first 1.5s of the rest interval and the 2.5s of the action interval were considered for each trial (Figure \ref{fig:experim-proced}). 
Subsequently, multiclass classification considering the four words (\textit{left}, \textit{right}, \textit{up} and \textit{down}) of inner speech was carried out. Action interval 2.5s were used for each trial (Figure \ref{fig:experim-proced}).

The following ML analyses were performed for both classification tasks.\\
Power Spectral Density (PSD) was used as a feature extraction technique before proceeding with the classification.
PSD was calculated using Welch's method and based on relative power in specific frequency bands: alpha (8-13 Hz), beta (13-30 Hz) and gamma (30-100 Hz). \\
The models were trained and tested on each subject individually, using K-fold cross validation. In this study, the data was split into four folds, resulting in a number of trials ranging from 237 to 285 trials (depending on the subject) in test set and from 713 to 855 trials in the training set in binary classification, while in multiclass classification from 118 to 142 trials in test set and from 357 to 428 trials in the training set.
The SVM and XGBoost classifiers were trained on PSD features vector with dimension $(n\_epochs, n\_channels * n\_band\_freqs)$ for each subject.\\
Since the number of features is too high compared to the number of trials of each subject, three possible solutions were analysed:
\begin{itemize}
	\item to apply Principal Component Analysis (PCA);
	\item to extract the most important features identified with the XGBoost classifier, both considering the subjects individually and making an intersection of the most important features in common to all subjects;
	\item to choose a subset of meaningful electrodes, since the neural correlates of inner speech processing are reported to be mainly present in the left hemisphere (see Section \ref{sec:introduction}).
\end{itemize}

The analysis carried out in the binary classification showed a difference between the action interval and the rest interval. It was therefore verified whether this difference could be associated with one or more time windows, in order to identify a particularly significant area in which the inner speech activity could be encoded.
The action interval has been split into 0.5s wide sliding windows with a 50\% overlap. For each window of the action interval, a binary SVM model was trained, considering all the resting state and using the features extracted with XGBoost.
The idea was to identify the best window for binary classification and then use it in multiclass classification.
Again a subject-based approach was used.

The analyzes performed do not justify the choice of one interval over another and this suggests continuing to consider the entire interval in the following tests carried out with deep learning methods. 

\subsection{Deep Learning approaches}
The DL models were trained and tested for the multiclass classification task on each subject individually, using nested cross-validation. \\
Three different types of analysis were performed in order to obtain the input data to train the models:
\begin{itemize}
	\item The PSD using the Welch’s method was calculated and the most important features were extracted using the XGBoost features importance vector.
	\item The raw data considering all the channels were used.
	\item The raw data considering only the channels associated with the most important features extracted using the XGBoost were used.
\end{itemize}
LSTM and BiLSTM networks were trained for each type of input data.

The \textit{Imagine Speech} dataset~\cite{coretto2017open} was chosen to confirm the validity of the model that obtained the best results on the \textit{Thinking Out Loud} dataset.

\section{Results and discussion}\label{sec:results-and-discussion}

\subsection{Machine Learning Models results}
In the binary classification task, among the various tests performed using the PCA, the best results for SVM were obtained without PCA and an accuracy of 79\% is obtained, while with XGBoost an accuracy of 81\% is obtained using PCA and explaining 99\% of the variance. 
Using the most important features extracted with XGBoost considering each subject individually, the SVM performances improve up to 80\% accuracy with a gain of 0.9.

In the multiclass classification task, the results obtained with SVM and XGBoost are very similar, respectively 26.2\% and 27.9\% accuracy. In this case, using the most important features extracted with XGBoost both considering each subject individually and in common to all subjects, the results are approximately the same. This means that the subjects have common characteristics relevant for classification. Furthermore, there are no particular differences when using all channels or only those of the left hemisphere.

\begin{figure}[]
	\centering
	\includegraphics[width=0.6\linewidth]{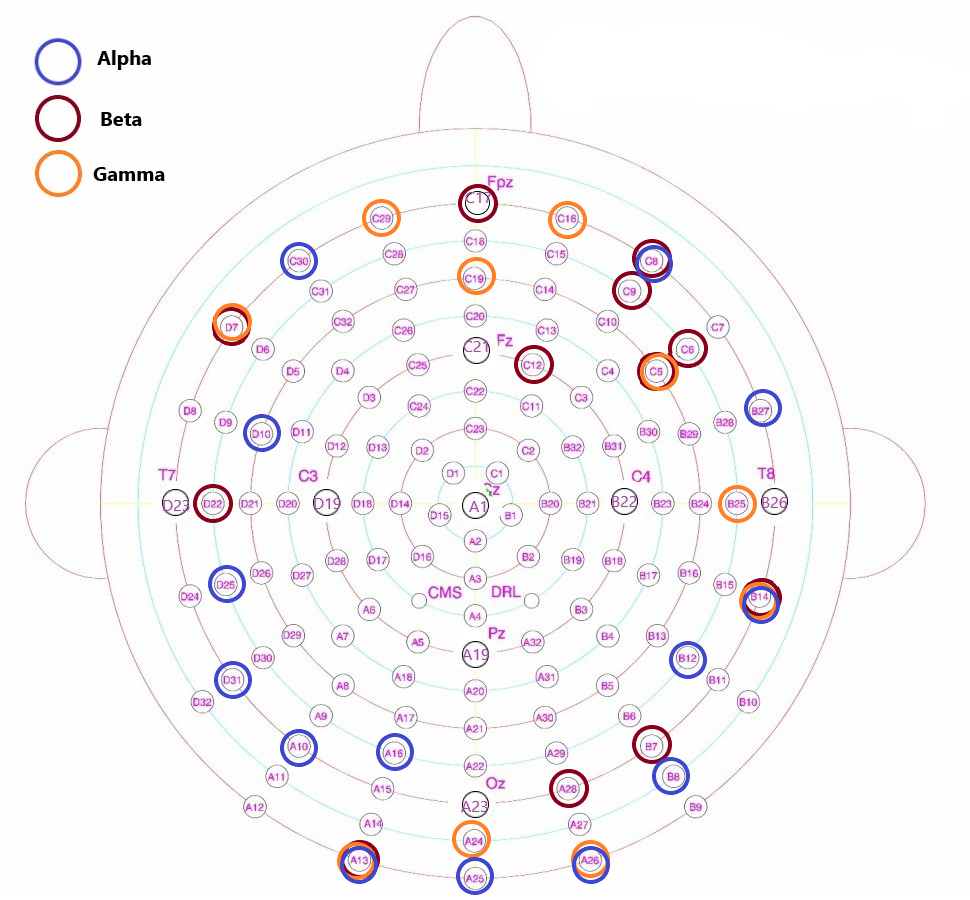}
	\caption{Features in common to all subjects in the multiclass task identified with XGBoost using all channels and gain=0.95}
	\label{fig:multi-clas-MIF-com-elec}
\end{figure}

The characteristics extracted in common to all subjects were analyzed and are highlighted in the Figure \ref{fig:multi-clas-MIF-com-elec}. 
Considering all the channels, the most affected areas are the occipital one, probably involved due to the visual signals presented on the screen, the temporal and the frontal one. Both the right and left hemispheres are involved. Since, even using only the electrodes identified in the left hemisphere, the performance still remains good, this could mean that there are electrodes in the left channels that compensate for the absence of the right ones. The frequency band most involved is alpha, usually associated with intense mental activity.
These features were used later for some analyzes carried out with the DL models.


\subsection{Deep Learning Models results}

This paragraph summarizes the performances obtained with the different deep learning models using the \textit{Thinking out loud dataset}. In table \ref{tab:dl-comparison} the results are shown averaging over all subjects.

\begin{table}[h]
	\centering
	\resizebox{\textwidth}{!}{%
		\begin{tabular}{|c|c|c|}
			\hline
			\textbf{Input Type} & \textbf{LSTM Accuracy} & \textbf{BiLSTM Accuracy} \\ \hline
			Most Important Features & 30.4 \% & 31.3\% \\ \hline
			Raw data (all channels) & 27.2\% & 36.1\% \\ \hline
			Raw data (channels Most Important Features) & 26.7\% & 33.1\% \\ \hline
		\end{tabular}%
	}
	\caption{Deep Learning Models comparison}
	\label{tab:dl-comparison}
\end{table}

In general, with BiLSTM the performances increases rather than using LSTM. This 
is due to the fact that BiLSTM is able to capture the sequential dependencies between data in both directions. The greatest improvement is obtained using raw data from all channels as input.

Looking at the average performance subject by subject, we have a repeating trend. Using both different models and different types of input data, the data of some subjects are classified better than others. Specifically, subjects 4 and 5 achieve an average performance which is always slightly lower. This could probably be related to the data acquisition phase of these volunteers. Maybe the placement of the electrodes was not perfect or their data is noisier. Analyzing the results of attention monitoring there are no differences with the other subjects, so the lack of attention in carrying out the task should not be the cause of the lower performance.

\begin{figure}[!ht]
	\centering
	\includegraphics[width=0.9\linewidth]{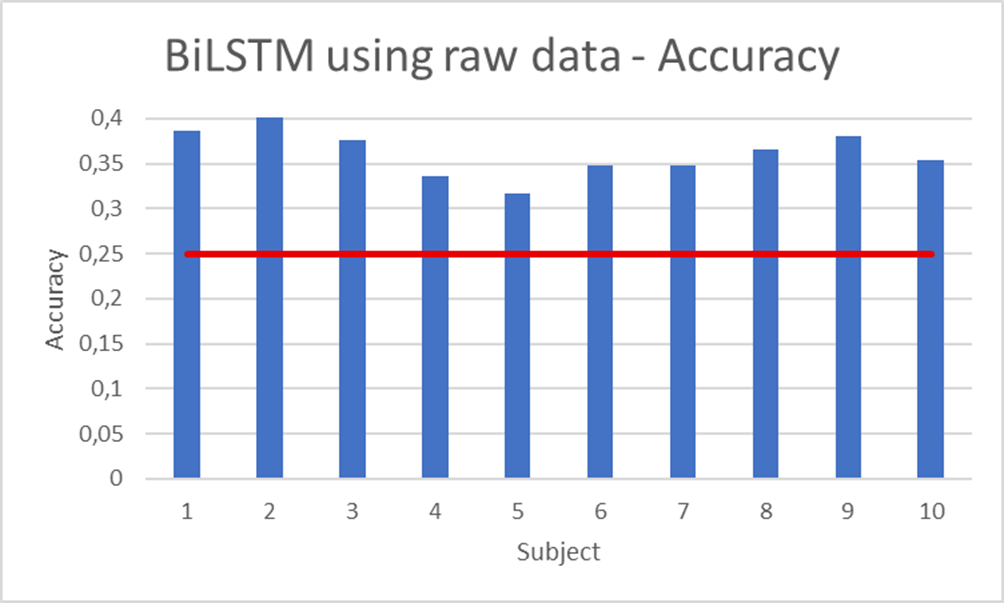}
	\caption{Accuracy of the BiLSTM network for multiclass classification using the \textit{Thinking out loud} raw data. The red line represents the chance level (25\%).}
	\label{fig:bilstm-raw-acc}
\end{figure}

The best model network is composed of a BiLSTM layer followed by two dense layer (with ReLU activation function) and a dense output layer (with softmax activation function). Two dropout layers were used to reduce overfitting. SGD was used as optimization method and categorical cross entropy as loss function.
Figure \ref{fig:bilstm-raw-acc} shown the results obtained for each subject. As we can see, all subjects achieve an accuracy above randomness (represented by the red line - 25\%). The mean accuracy considering all the subjects is 36.1\%.
Table \ref{tab:blstm-inner-acc-our} summarizes the results of our best model for each subject.

\begin{table}[!ht]
	\centering
	\begin{tabular}{|c|c|c|c|c|}
		\hline
		\textbf{Subject} & \textbf{Accuracy} & \textbf{Precision} & \textbf{Recall} & \textbf{F1-score} \\ \hline
		\textbf{Sub 1} & 38.60\% & 40.37\% & 38.93\% & 37.87\% \\ \hline
		\textbf{Sub 2} & 40.17\% & 40.06\% & 40.43\% & 39.56\%  \\ \hline
		\textbf{Sub 3} & 37.60\% & 38.25\% & 37.50\% & 35.50\% \\ \hline
		\textbf{Sub 4} & 33.67\% & 34.44\% & 33.69\% & 32.81\% \\ \hline
		\textbf{Sub 5} & 31.67\% & 32.13\% & 31.94\% & 31.06\% \\ \hline
		\textbf{Sub 6} & 34.81\% & 37.69\% & 33.00\% & 33.75\% \\ \hline
		\textbf{Sub 7} & 34.83\% & 36.00\% & 34.94\% & 34.31\% \\ \hline
		\textbf{Sub 8} & 36.60\% & 37.69\% & 36.88\% & 34.88\% \\ \hline
		\textbf{Sub 9} & 38.00\% & 38.19\% & 37.75\% & 37.31\% \\ \hline
		\textbf{Sub 10} & 35.33\% & 34.50\% & 34.94\% & 34.38\% \\ \hline
		\textbf{Average} & 36.12\% & 36.93\% & 36.00\% & 35.14\% \\ \hline
	\end{tabular}
	\caption{BiLSTM performance for each subject on the 4-class inner speech classification task using the \textit{Thinking out loud} raw data}
	\label{tab:blstm-inner-acc-our}
\end{table}

Table \ref{tab:results-thinking-out-loud} shows a comparison of our proposed approaches and works in the literature using the \textit{Thinking out loud} dataset.

\begin{table}[]
	\centering
	\begin{tabular}{|c|c|c|}
		\hline
		\textbf{Classifier} & \textbf{Input Data} & \textbf{Accuracy} \\ \hline
		SVM & PSD Features (channels left hemisphere) + PCA (0.99) & 26.2\% \\ \hline
		XGBoost & PSD Features + PCA (0.99) & 27.9\% \\ \hline
		\multirow{3}{*}{LSTM} & Most Important Features & 30.4\% \\ \cline{2-3} 
		& Raw data (all channels) & 27.2\% \\ \cline{2-3} 
		& Raw data (channels Most Important Features) & 26.7\% \\ \hline
		\multirow{3}{*}{BiLSTM} & Most Important Features & 31.3\% \\ \cline{2-3} 
		& Raw data (all channels) & \textbf{36.1\%} \\ \cline{2-3} 
		& Raw data (channels Most Important Features) & 33.1\% \\ \hline \hline
		EEGNet & Raw Data (channels left hemisphere) & 29.67\% \cite{van2021inner}\\ \hline
	\end{tabular}
	\caption{Comparison of our ML and DL models with results in the literature using the \textit{Thinking Out Loud} dataset.}
	\label{tab:results-thinking-out-loud}
\end{table}

\begin{figure}[!ht]
	\centering
	\includegraphics[width=0.9\linewidth]{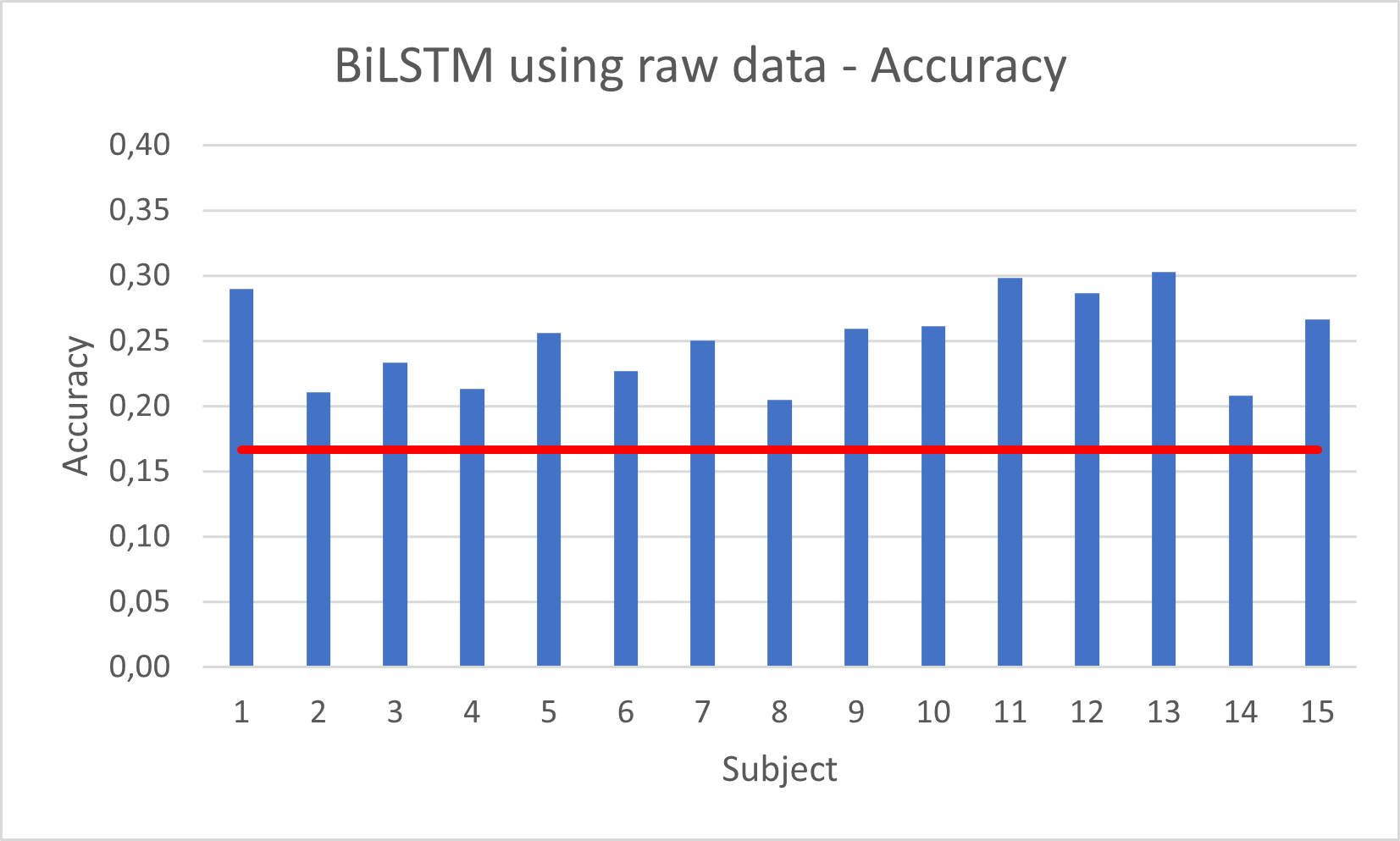}
	\caption{Accuracy of the BiLSTM network for multiclass classification using the \textit{Imagined Speech} raw data. The red line represents the chance level (16.7\%).}
	\label{fig:bilstm-raw-acc_d2}
\end{figure}

The \textit{Imagined Speech} dataset was chosen to confirm the validity of the model that obtained the best results on the \textit{Thinking Out Loud} dataset.
Figure \ref{fig:bilstm-raw-acc_d2} shown the results obtained for each subject. As we can see, all subjects achieve an accuracy above chance (represented by the red line - 16.67\%). The mean accuracy considering all the subjects is 25.1\%.

Table \ref{tab:results-imagined-speech} shows a comparison of our proposed approach and works in the literature using the \textit{Imagined Speech} dataset.

\begin{table}[]
	\centering
	\begin{tabular}{|c|c|c|}
		\hline
		\textbf{Classifier} & \textbf{Input Data} & \textbf{Accuracy} \\ \hline
		BiLSTM & Raw data (all channels) & \textbf{25.1\%} \\ \hline \hline
		RF & Relative Wavelet Energy (RWE) & 18.58\% \cite{coretto2017open}\\ \hline \hline
		EEGNet & Raw Data (all channels) & 24.97\% \cite{cooney2020evaluation}\\ \hline
	\end{tabular}
	\caption{Comparison of our DL model with results in the literature using the \textit{Imagined Speech} dataset}
	\label{tab:results-imagined-speech}
\end{table}

\section{Conclusions}\label{sec:conclusion}

Inner speech recognition decoding EEG signal is still an open field of research. 
Few datasets are available in the literature and the classification performance, even if above chance, is still very low. 
The results obtained with this work confirm that the adoption of BiLSTM architecture increases the performance of classification with respect to those in the state-of-the-art. In particular, the model designed for the \textit{Thinking Out Loud} dataset was tested on a different one, acquired using a similar experimental protocol, confirming the validity of our proposal. 
The best classifier is obtained considering raw data of all channels, denoting that a deeper analysis on the most significant features should be performed, recalling that inner speech recognition should be considered in BCI applications where classification should be performed in real time.
Finally, the need of more numerous and less noisy datasets is crucial for further development in this field of research.

\bibliographystyle{splncs04}
\bibliography{mybib}


\end{document}